\documentstyle[12pt]{article}
\def\lsim{\mathrel{\rlap{\lower4pt\hbox{\hskip1pt$\sim$}}
    \raise1pt\hbox{$<$}}}         
\def\gsim{\mathrel{\rlap{\lower4pt\hbox{\hskip1pt$\sim$}}
    \raise1pt\hbox{$>$}}}         

\def\overleftrightarrow#1{\vbox{\ialign{##\crcr
    $\leftrightarrow$\crcr
    \noalign{\kern 1pt\nointerlineskip}
    $\hfil\displaystyle{#1}\hfil$\crcr}}}

\begin{document}
\begin{center}

{\bf QCD Sum Rules and the Induced Pseudoscalar Coupling}
\vspace{.5in}

E.M. Henley \\

{\em Department of Physics and Institute for Nuclear Theory, \\
Box 351560, University of Washington, Seattle, Washington 98195}

\vspace{2mm}
W-Y. P. Hwang \\

{\em Asia-Pacific and Leung Center for Cosmology and Particle Physics,\\
Center of Theoretical Sciences, \\
Department of Physics, and Institute of Astrophysics,\\
National Taiwan University, Taipei 10617, Taiwan}

\vspace{2mm}
L.S. Kisslinger \\
{\em Department of Physics, Carnegie-Mellon University, \\
Pittsburgh, Pennsylvania 15213}

\vspace{.5in}
{\bf Abstract}
\end{center}

We present an extension of the QCD sum rule method in the external fields so as to
determine the induced pseudoscalar coupling constant $g_P$, which tests the
validity of the partially conserved axial current (PCAC) hypothesis. This is
essentially that we pick out the "higher-order" effects of both the hadron and
quark (QCD) sides. A specific QCD sum rules for $g_P$ is
obtained and its prediction is briefly analyzed. It turns out that the final
prediction on $g_P$ is extremely stable. In view of the versatile nature of the
present QCD sum rule methods, we appendix some discussions on the possible future
of the method.

\vspace{.5in}

\noindent
PACS Indices: 11.40.Ha, 12.38.Lg, 11.30.Rd, 11.50.Li

\newpage


\section{Introduction}

The matrix element of the isovector axial current between the on-shell nucleon
states,$^1$ with $q_\mu \equiv p_\mu -p'_\mu$,
\begin{eqnarray}
&&<n(p')\mid A_\mu(0) \mid p(p)> \nonumber\\
&=& {\bar u}(p') \{ f_A(q^2)\gamma_\mu \gamma_5 + f_P(q^2) q_\mu \gamma_5 \}
u(p),
\label{eq:matrix1}
\end{eqnarray}
plays a fundamental role in the description of semileptonic weak interactions
such as beta decay, muon capture, and neutrino-induced charged weak reactions. Here
$g_A \equiv f_A(q^2=0) = 1.2695 \pm 0.0029$ is the axial coupling$^2$ while
$g_P \equiv f_P(q^2=0)$ is the induced pseudoscalar coupling. It is well-known
that the one-pion pole contributes to the induced pseudoscalar
form factor $f_P(q^2)$, in the way that the proton turns into the neutron
by emitting a pion which in turn couples to $W^\pm$. Such a one-pion pole
contribution is$^3$
\begin{eqnarray}
f_P(q^2) = {\sqrt 2} g_{\pi NN}(q^2) \cdot {1\over q^2 - m_\pi^2}\cdot
{\sqrt 2} f_\pi (q^2),
\label{eq:fp2}
\end{eqnarray}
where $f_\pi(q^2=0)$ is the pion decay constant and
$g_{\pi NN}(q^2)$ the strong $\pi NN$ coupling. Assuming that this is the
only major contribution to $f_P(q^2)$ and using the partially conserved
axial current (PCAC) hypothesis that $\partial_\mu A^\mu (x) \sim O(m_\pi^2)
\sim 0$, we find, in the limit that $m_\pi^2=0$,
\begin{eqnarray}
&  M f_A(q^2) - g_{\pi NN}(q^2) f_\pi(q^2) \approx 0;\nonumber \\
&  M g_A \approx g_{\pi NN} f_\pi,
\label{eq:gt3}
\end{eqnarray}
where $M \equiv {1\over 2}(m_n +m_p)$ is the nucleon mass and $m_\pi$ is the
pion mass. Note that smooth extrapolation to $q^2=0$ is needed in obtaining the
Goldberger-Treiman (GT) relation,$^4$ the second identity in Eq. (3).

\section{Formulation}

Although the method of QCD sum rules as originally developed$^5$ was applied
to the study of hadronic properties in the region of about $1\, GeV$, Ioffe and
Smilga$^6$ developed techniques for embedding hadrons in an external field in
order to derive static properties in terms of the condensates, including
induced condensates which introduce new parameters.
The method of QCD sum rules in the presence of external axial fields has been
employed$^{7,8,9}$ to extract the axial coupling $g_A$. It is the purpose of
the present paper to indicate how this method may be suitably generalized to
obtain the induced pseudoscalar coupling $g_P$.

We begin by briefly reviewing the method for an external axial field.
The starting point is the polarization function in an external axial
field, which we call $Z_\mu$.  The correlation function, $\Pi(p)$, is defined
as$^{6-9}$
\begin{eqnarray}
\Pi(p) \equiv i \int d^4x e^{ip\cdot x} <0\mid T(\eta(x) {\bar \eta}(0))
\mid 0>,
\label{eq:corr4}
\end{eqnarray}
where for the nucleon current we may use a standard form
\begin{eqnarray}
\eta(x) &= \epsilon^{abc} \{ u^a(x)^T C \gamma_\mu u^b(x)\} \gamma^\mu
\gamma^5 d^c (x), \nonumber\\
&<0\mid \eta(0) \mid N(p)> \equiv \lambda_N v_N(p),
\label{eq:eta5}
\end{eqnarray}
with $C$ the charge conjugation operator, $a,b,c$ color indices, and
$v_N(p)$ Dirac spinor for the nucleon normalized such that ${\bar v}(p) v(p)= 2
M$. Embedding the system in an external $Z_\mu$ field and introducing
intermediate states, one can express the correlation function in the limit of
a constant external field, $Z_\mu(x) = Z_\mu$, as$^9$

\begin{eqnarray}
\Pi(p) = - \mid \lambda_N \mid^2 {1\over {\hat p} -M_N} g_A {\hat Z}
\gamma_5 {1\over {\hat p} -M_N} + \cdot\cdot\cdot,
\label{eq:rhs6}
\end{eqnarray}
with ${\hat a} \equiv \gamma_\mu a^\mu$, where Eq. (1) has
been adopted. The term shown in Eq. (6) corresponds to nucleon
intermediate states, while the continuum contributions to $\Pi$ are implied by
the ellipses in Eq. (6).  Eq. (6) is the expression for the
phenomenological form, in which $\Pi(p)$ is evaluated at the baryon level.
When evaluating the correlation function $\Pi(p)$ at the quark level and
comparing it with Eq. (6), one is led to three sum rules involving
$g_A$, which may not be consistent among themselves although there is indeed
one sum rule$^9$ which seems most appropriate for $g_A$.

   We note that Eq. (1) gives the on-shell matrix element of
the axial current. In treating the correlator $\Pi(p)$ in the presence of an
external axial field $Z_\mu(x)$, one may consider the slightly off-shell
nucleon matrix elements, where additional off-shell form factors can occur. For
instance,
\begin{eqnarray}
&&<N(p',\lambda')\mid J^5_\mu(0) \mid N(p,\lambda)>\nonumber\\
&=&{\bar u}_{\lambda'}(p')\{ G_1(q^2) \gamma_\mu \gamma_5 +G_2(q^2) q_\mu
\gamma_5 + G_3(q^2) i \sigma_{\mu\nu}P^\nu \gamma_5\} u_\lambda(p),
\label{eq:matrix7}
\end{eqnarray}
with $P_\mu \equiv p'_\mu +p_\mu$. In the on-shell limit, this reduces to
Eq. (1) with $g_A = G_1 + 2 M G_3$ and $g_P = G_2 - G_3$.
Among the three sum rules which one obtains by comparing coefficients of $p
\cdot Z {\hat p} \gamma_5$, ${\hat Z} \gamma_5$, and $i \sigma_{\mu\nu}Z^\mu
p^\nu \gamma_5$, only the axial coupling $g_A$ enters in the on-shell limit,
making it difficult to determine the pseudoscalar coupling $g_P$. This is a
general problem for obtaining the induced couplings [such as the anomalous
magnetic moment and the pseudoscalar coupling] or form factors [such as the
$q^2-$dependence of $f_A(q^2)$] in the QCD sum rule method.

   To obtain a QCD sum rule for the induced pseudoscalar coupling $g_P$, we
consider the external axial field $Z_\mu(x)$ as follows:
\begin{eqnarray}
Z_\mu(x) = Z_\mu^0 + {1\over 2} Z_{\mu\nu} x^\nu,
\label{eq:z8}
\end{eqnarray}
where $Z_\mu^0$ and $Z_{\mu\nu}$ are constants. We have, in momentum space,
\begin{eqnarray}
Z_\mu(q) = Z_\mu^0\delta^4(q) -{i\over 2}Z_{\mu\nu}\partial^\nu\delta^4(q).
\label{eq:z9}
\end{eqnarray}
We shall focus our attention on the $Z_{\mu\nu}$ terms, in a way
similar to the work of calculating the anomalous magnetic moments$^6$.
This implies that Eq. (6) is to be replaced by
\begin{eqnarray}
\Pi_A(p) &= {\mid \lambda_N \mid^2 \over (p^2 -M^2)^2}
\{ {1\over 2}Z^{\mu\nu}_A \sigma_{\mu\nu}\gamma_5 [M G_1 +(p^2 +M^2) G_3]
&\nonumber\\
&\qquad + {1\over 2}Z^{\mu\nu}_A \sigma_{\mu\nu}\gamma_5 {\hat p} [G_1 + 2M G_3]
&\nonumber\\
&\qquad + i Z^{\mu\nu}_A \gamma_\mu p_\nu\gamma_5 {1\over p^2 -M^2}[(p^2 +M^2)
G_1 + 4 Mp^2 G_3] &\nonumber\\
&\qquad + i Z^{\mu\nu}_A \gamma_\mu p_\nu\gamma_5 {\hat p}{1\over p^2-M^2}[2 M
G_1  +2 (p^2 +M^2) G_3] \}
+ \cdot\cdot\cdot, &
\label{eq:rhs10}
\end{eqnarray}
for the antisymmetric part of $Z_{\mu\nu}$, and

\begin{eqnarray}
\Pi_S(p) &= {\mid \lambda_N \mid^2 \over (p^2 -M^2)^2}
\{ -{i\over 2} g_{\mu\nu} Z^{\mu\nu}_S \gamma_5 [M G_1 + 2 p^2 G_3
+ (-p^2 +M^2) G_2]
&\nonumber\\
&\qquad - {i\over 2} g_{\mu\nu} Z^{\mu\nu}_S \gamma_5 {\hat p} [G_1 + 2M G_3]
&\nonumber\\
&\qquad + i Z^{\mu\nu}_S p_\mu p_\nu\gamma_5 {1\over p^2 -M^2}[ 2M G_1 + 2(p^2
+M^2) G_3] &\nonumber\\
&\qquad + i Z^{\mu\nu}_S p_\mu p_\nu\gamma_5 {\hat p} {1\over p^2 -M^2}[ 2 G_1
+ 4 M G_3] &\nonumber\\
&\qquad + i Z^{\mu\nu}_S \gamma_\mu p_\nu\gamma_5 {1\over p^2-M^2}[(p^2 + M^2)
G_1 + (3 p^2 +M^2) M G_3]&\nonumber\\
&\qquad + i Z^{\mu\nu}_S \gamma_\mu p_\nu\gamma_5 {\hat p} {1\over p^2-M^2}
[2 M G_1 + (p^2 + 3M^2) G_3] \}
+ \cdot\cdot\cdot, &
\label{eq:rhs11}
\end{eqnarray}
for the symmetric part of $Z_{\mu\nu}$, where we have $Z^{\mu\nu} \equiv
Z^{\mu\nu}_A + Z^{\mu\nu}_S$ with $Z^{\mu\nu}_A = -Z^{\nu\mu}_A$ and
$Z^{\mu\nu}_S  =Z^{\nu\mu}_S$.

   We note that, in the on-shell limit ($p^2 \to M^2$), all the coefficients in
Eqs. (10) and (11) reduce to $g_A$ ($=G_1 +2MG_3$) {\it except} the one
proportional to $g_{\mu\nu} Z^{\mu\nu}_S \gamma_5$ (in which the induced
pseudoscalar coupling $f_P$ or $G_2$ enters). Therefore, we wish to focus on
the sum rule obtained by working with this specific Lorentz structure.

   Next, we need to evaluate the correlation function at the quark level,
making use of the quark propagator in the presence of gluonic and $Z$ fields.
The quark propagator is defined by
\begin{eqnarray}
i S^{ab}_{ij}(x) \equiv <0\mid T(q^a_i(x) {\bar q}^b_j(0))\mid 0>.
\label{eq:quark12}
\end{eqnarray}
Following the method of Ref. 6, including terms up to second order in the
Taylor expansion, we find, in the presence of $Z_\mu^0$ and $Z_{\mu\nu}^S$,

\begin{eqnarray}
i S^{ab}(x) &= &{\delta^{ab}\over (2\pi)^4 x^4} \{i{\hat x} - g
(x\cdot Z^0 -{1\over 4} Z^{\mu\nu}_S x_\mu x_\nu) {\hat x} \gamma_5\}
\nonumber\\
&& + {i\over 32 \pi^2 x^2} g_c {\lambda^n_{ab}\over 2} G^n_{\mu\nu}
({\hat x} \sigma^{\mu\nu} +\sigma^{\mu\nu}{\hat x})\nonumber\\
&&+\delta^{ab} <{\bar q} q>
\{ -{1\over 12}(1 +{1\over 16} x^2 m_0^2) + {1\over 12} g \chi
{\hat Z} \gamma_5 -{1\over 12} g \chi' g_{\mu\nu}Z^{\mu\nu}_S \gamma_5
\nonumber\\
&&\qquad - {1\over 36} g \sigma^{\mu\nu} x_\mu Z^0_\nu \gamma_5
+ {1\over 216}g \kappa ({5\over 2} x^2 {\hat Z} - x\cdot Z {\hat x})\gamma_5
\nonumber\\
&&\qquad+
{1\over 192} g \kappa' g_{\mu\nu}Z^{\mu\nu}_S x^2 \gamma_5\} + \cdot\cdot\cdot.
\label{eq:quark13}
\end{eqnarray}
Here the condensate parameters are defined by
\begin{eqnarray}
&<0\mid {\bar q} g_c \sigma\cdot G q \mid 0> = - m^2_0 <{\bar q} q>,\nonumber\\
&<0\mid {\bar q} g_c {\tilde G}_{\mu\nu}\gamma^\nu q \mid 0> = g \kappa
Z_\mu^0 <{\bar q} q>,\nonumber\\
&<0\mid {\bar q} i g_c \sigma^{\mu\nu}G_{\mu\nu} \gamma_5 q \mid 0> = g
\kappa' g_{\mu\nu} Z^{\mu\nu}_S <{\bar q} q>, \nonumber\\
&<0\mid {\bar q} \gamma_\mu \gamma_5 q \mid 0> = g\chi Z_\mu^0<{\bar q} q>,
\nonumber\\
&<0\mid {\bar q} i\gamma_5 q \mid 0> = g\chi' g_{\mu\nu} Z^{\mu\nu}<{\bar q} q>,
\label{eq:cond14}
\end{eqnarray}
Our form for the quark propagator, Eq. (13), is the same as that of Ref. 9
except for the additional terms related to $Z^{\mu\nu}_S$ $-$ these new terms
resemble the terms in $Z_\mu^0$ (and if so desired) may be represented
pictorially by the same diagrams as shown in Ref. 9. Note that the first four
terms in Eq. (13) are the perturbative free quark propagator, while
the remaining are nonperturbative terms, proportional to the quark condensate
$<{\bar q}q>$. The other quantities
appearing in Eq. (13) are the $Z$-quark coupling constant [$g =
g_u = -g_d$ for the isovector axial coupling $g_A$ or $g= g_u = g_d$ for the
isoscalar axial coupling $g_A^S$].

   We may proceed to evaluate the correlation function $\Pi(p)$ at the quark
level by considering the processes up to a certain dimension. This has become a
routine but standard exercise in the QCD sum rule practices.$^{6,8,9}$ In
particular, we use
\begin{eqnarray}
<0\mid T(\eta(x) {\bar \eta}(0))\mid 0> &=& -2\epsilon^{abc}\epsilon^{a'b'c'}
Tr\{ i S(x)_u^{bb'} \gamma_\nu C i S(x)_u^{aa'T} C \gamma_\mu\}\nonumber\\
&& \cdot \gamma_5 \gamma^\mu i S(x)_d^{cc'} \gamma^\nu \gamma_5.
\label{eq:corr15}
\end{eqnarray}
In the present case, we obtain, up to dimension $D=6$ (as compared to the
leading diagram, counted as $D=0$),
\begin{eqnarray}
&+{24\over \pi^6}{{\hat x}\over x^{10}}g_d (-{1\over 4}
Z^{\mu\nu}_S x_\mu x_\nu)\gamma_5
+ {2\over\pi^4 x^6} g_d \chi' <{\bar q}q>
g_{\mu\nu} Z^{\mu\nu}_S \gamma_5
&\nonumber\\
& -{1\over 32\pi^6} {{\hat x}\over x^6} g_u
<g_c^2 G^2> (-{1\over 4} Z^{\mu\nu}_S x_\mu x_\nu)\gamma_5
 -{1\over 8\pi^4 x^4} g_d \kappa' <{\bar q} q>
g_{\mu\nu} Z^{\mu\nu}_S \gamma_5
&\nonumber\\
& -{1\over 3} {{\hat x}\over \pi^2 x^4} g_d
<{\bar q} q>^2 (-{1\over 4} Z^{\mu\nu}_S x_\mu x_\nu)\gamma_5,&
\label{eq:lhs16}
\end{eqnarray}
which, upon Fourier transform, gives rise to the correlation
function $\Pi(p)$ evaluated at the quark level, the left-hand-side (l.h.s.)
of the sum rule. Comparing the coefficients for both the expressions
$i g_{\mu\nu} Z^{\mu\nu}_S \gamma_5$ and $i Z^{\mu\nu}_S \gamma_\mu p_\nu
\gamma_5$ (and thus obtaining two QCD sum rules soon to be combined),
performing Borel transform on both the r.h.s. and l.h.s., taking into account
the anomalous dimensions, and combining the two sum rules, we arrive at the
following QCD sum rule,
\begin{eqnarray}
&& - g_d{M_B^6\over 8} L^{-4/9}E_2 + {g_u <g_c^2 G^2>M_B^2\over 8} L^{-4/9}E_0
- {g_d M_B^4\chi' a\over 2} M L^{-4/9}E_1 -g_d {2\over 3} a^2 L^{4/9}
\nonumber\\
&&\qquad - {g_d M_B^2\kappa'a\over 4 L^{68/81}} M E_0
= \beta^2_N\;e^{-M^2/M^2_B} (g_A + M g_P),
\label{eq:sumgp17}
\end{eqnarray}
with $a = -(2\pi)^2 <{\bar q}q>$ and $L = 0.621 ln(10 M_B)$, corresponding to
$\Lambda_{QCD} = 0.1 \, GeV$ with the Borel mass, $M_B$, in $GeV$ and
$\beta_N^2 \equiv (2\pi)^4 \lambda_N^2/4$ ($\approx 0.26\,GeV^6$). Note that
the factors $E_0 = 1- e^{-x}$, $E_1 = 1- (1+x)
e^{-x}$, and $E_2 = 1- (1+x+{1\over 2} x^2)e^{-x}$, with $x\equiv
W^2/M_B^2 \approx (2.3 \, GeV^2)/M_B^2$ (see Ref. 9), describe the
contributions from the excited states through perturbative QCD
method$^{10,11}$.

We recall the QCD sum rule for $g_A$ as obtained from Ref. 9, again up to
dimension $D=6$,
\begin{eqnarray}
{M^4_B E_2\over 8 L^{4/9}} &+&{1\over 32 L^{4/9}}<g_c^2 G^2> E_0
- {1\over 18 L^{68/81}}\kappa a E_0+{5\over 18 M^2_B} a^2 L^{4/9} \nonumber\\
& = &  \beta_N^2 exp (-M^2/M^2_B) g_A.
\label{eq:sumga18}
\end{eqnarray}
Combining Eqs. (18) and (17) [the latter with $g_u=-g_d =1$], we obtain the
QCD sum rule for $g_P$:
\begin{eqnarray}
&& {M_B^4\chi' a\over 2} M L^{-4/9}E_1 + { M_B^2E_0\over L^{68/81}}({M\kappa'a
\over 4} +{\kappa a\over 18}) + {3 <g_c^2 G^2>M_B^2\over 32} L^{-4/9}E_0
\nonumber\\
&&\quad + {7\over 18} a^2 L^{4/9} = \beta^2_N\;e^{-M^2/M^2_B} M g_P,
\label{eq:sumgp19}
\end{eqnarray}
which is the main result of this paper.

It is of interest to derive a similar QCD sum rule for the {\it isoscalar}
pseudoscalar coupling $g_P^S$. Along the same line [as from Eq. (17) up to Eq.
(19)], we obtain [with $g_u= g_d =1$]
\begin{eqnarray}
&& -{M_B^4 E_1 \over L^{4/9}} ({M\chi' a\over 2} +{\chi a\over 6}) - { M_B^2
E_0\over L^{68/81}}({M\kappa'a\over 4} - {\kappa a\over 18})
+ {3 <g_c^2 G^2>M_B^2\over 32} L^{-4/9}E_0
\nonumber\\
&&\quad + {1\over 18} a^2 L^{4/9} = \beta^2_N\;e^{-M^2/M^2_B} M g_P^S.
\label{eq:sumgps20}
\end{eqnarray}
It is known that, in the absence of the pion-pole dominance for the isoscalar
channel, $g_P^S$ is small, signaling the cancellation ${M\chi' a\over 2}
+{\chi a\over 6}\approx 0$ (since the other contribution is numerically small).
However, it should be kept in mind that the susceptibilities in the isoscalar
channel, such as $\chi^S$ in Eq. (20), may differ significantly from those in
the isovector channel, i.e. those for Eq. (19), where Goldstone pions play
a very important role. Indeed, it is reasonable to expect that the
induced condensate $<0\mid {\bar q} i\tau_j\gamma_5 q \mid 0>\mid_{Z_\mu\nu}$
($\equiv 2 g\chi'g_{\mu\nu}Z^{\mu\nu}_j < {\bar q} q>$)
is related closely to $<0\mid {\bar q} i\gamma_5 q \mid 0>\mid_\pi$
($\equiv g_{\pi q} \chi_\pi \pi_j < {\bar q} q>$), resulting in
a susceptibilty$^{12}$:
\begin{equation}
2 g\chi' f_\pi \approx g_{\pi q}\chi_\pi \approx 8.9/a\, GeV^{-1},
\label{eq:pcac21}
\end{equation}
where $f_\pi$ ($=93\, MeV$) sets the scale for chiral symmetry breaking.
Without any reliable method to determine $\chi'$, this relation can only be
considered as an order-of-magnetic estimate.

Using the esitmates$^9$, $a\approx 0.55\, GeV^3$, $<g_c^2 G^2>\approx 0.47\,
GeV^4$, $\kappa a \approx 0.140\, GeV^4$, and ${M\kappa'a\over 4} +{\kappa
a\over 18}\approx 0$, we obtain, from the sum rule (19) evaluated at $M_B=1.1\,
GeV$,
\begin{equation}
g_P \approx (-132.4 \pm 5.7),
\label{eq:pcac22}
\end{equation}
where almost all of (-132.4) comes from the dominant $\chi'a$ contribution
(i.e. the first term) while the error bar comes from changing $M_B$ from
$1.1 GeV$ to $1.1 \pm 0.1 GeV$. The relatively unknown in
${Mk'a\over 4} + {ka\over 18}$ is small because the second term is clearly
known (to be small). In addition, this result is
vey stable with respect to the Borel mass $M_B$.

There are two aspects in connection with the experimental test of PCAC: The
first aspect has to do with the value of $g_A$, which should be in accord with
the GT relation$^4$, the second identity in Eq. (3). The second aspect has to
do with the value of $g_P$ which, according to Eqs. (2) and (3), reads
\begin{equation}
f_P(q^2) \approx {2 M f_A(q^2)\over q^2-m_\pi^2}, \qquad g_P \approx
-{2Mg_A\over m_\pi^2} \approx -124.
\label{eq:pcac23}
\end{equation}
The second aspect seems to be respected reasonably well by comparing it with
Eqs. (21) and (22), although it is obviously
desirable to obtain a quantitative treatment of the susceptibility $\chi'$.

As for the first aspect, we may begin with the pseudovector coupling for the
$\pi NN$ interaction,
\begin{equation}
{\cal L}_{\pi NN} = {f_{\pi NN}\over
 m_\pi}\;\bar\psi_N\;i\gamma^\mu\gamma^5\vec
\tau\cdot\psi_N\nabla_\mu\vec\phi_\pi.
\label{eq:s24}
\end{equation}
Accordingly, if we treat $\nabla_\mu\phi_\pi$ as a constant external axial
vector field, the resultant QCD sum rule for $f_{\pi NN}/m_\pi$
is identical to that$^9$ for $g_A$, except that, at the quark level, we have,
making use of the effective chiral quark theory$^{13, 14}$
\begin{equation}
{\cal L}_{\pi qq} = {1\over 2f_\pi}\;\bar\psi_q\;i\gamma^\mu
\gamma^5\vec\tau\psi_q\nabla_\mu\vec\phi_\pi\;,
\label{eq:s25}
\end{equation}
where $f_\pi$ is the pion decay constant. [For $g_A$, we begin with the
coupling at the quark level, $g_u=-g_d=1$.]  We thus obtain
\begin{eqnarray}
{f_{\pi NN}\over m_\pi} & = & {g_A\over 2f_\pi}\;,\nonumber \\
g_{\pi NN} & = & f_{\pi NN}\;{2M\over m_\pi} = {g_AM\over f_\pi},
\label{eq:s26}
\end{eqnarray}
which is just the Goldberger-Treiman (GT) relation$^4$. This proof
of the GT relation does not involve the pseudoscalar coupling $g_P$, the main
focus of this paper.

\section{Conclusion and Discussions}

To sum up, we have in this paper presented a suitable extension of the QCD sum
rule method which enables us to obtain a QCD sum rule for the
induced pseudoscalar coupling constant $g_P$, an entity of significance
for testing the validity of the partially conserved axial current (PCAC)
hypothesis.

Maybe we spend some paragraphs in discussing the overall future of the QCD
sum rule methods.

First of all, we have to go beyond the leading order in order to get
the induced pseudoscalar coupling $g_P$. The situation is similar to
the anomalous magnetic moments of baryons. One issue is whether we could
determine the newly condensate parameters, such as Eq. (14). The increase in
unknowns could be more than what we want to solve. For this, it is useful
to understand the theory more and to try to derive the relations among
these condensate parameters.$^{12,14}$

We have used the sum rule for $g_A$ [Eq. (18)] to obtain the sum rule
for $g_P$ [Eq. (19)]. In addition, it is assumed that the induced pseudoscalar
in the isoscalar channel is presumably very small, signaling that 
${M\chi'a\over 2}+{\chi a\over 6}$ and ${M\kappa'a\over 4}
- {\kappa a\over 18}$ be small. This is why we could make some numerical
prediction.

The value of $M_B$, ($=1.1 GeV$), is taken from the QCD sum rules for the nucleon
mass, for which the minimization has a meaning. Similarly, in bag models, minimization
to get the certain mass has a similar meaning. In contrast, the minimization in the
case of $g_A$, $g_P$, $\mu_P$, etc. should not take the fundamental meaning, as
compared to the mass or energy. Once the value of $M_B$ for nucleons is determined,
it should be used for predicting other fundamental parameters. This point should
always be emphasized for a QCD sum rule calculation.

In other words, QCD sum rule methods allow us to achieve the following: 
The basic properties of the nucleons are predicted using a universal
Borel mass $M_B$, based on the same set of the condensate parameters.
The adoption of the Borel transform is to improve the convergence of
the methods - it's not required nor necessary.

\vskip 0.5 true cm
\centerline{\bf Acknowledgements}
\bigskip
The work of W-Y.P. Hwang was supported in part by the National Science Council of
R.O.C. The work of E. M. Henley was supported in part by
the U.S. Department of Energy under grant DE-FG06-88ER40427, while that of
L.S. Kisslinger was supported in part by the National Science Foundation grant
PHY-9319641. This work was used to be supported by the N.S.C. of
R.O.C. and the National Science Foundation of U.S.A. as a cooperative
research project. W-Y.P. Hwang wishes to acknowledge the Institute for Nuclear
Theory for the hospitalities during his participation of the Program INT-96-1
(on ``Quark and Gluon Structure of Nucleons and Nuclei''), when part of this
manuscript was initially completed.

\vskip 0.5 true cm
\begin{enumerate}

\item For notations, see, e.g., T.-P. Cheng and L.-F. Li, {\it Gauge Theory of
Elementary Particle Physics} (Clarendon Press, Oxford, 1984).

\item Particle Data Group, {\it Review of Particle Physics} J. Phys. G: Nucl. Part.
Phys. {\bf 33}, 1(2006).

\item See, e.g., E.D. Commins and P.H. Bucksbaum, {\it Weak Interactions of
Leptons and Quarks} (Cambridge University Press, 1983), p. 169 $-$ 172.

\item
M. L. Goldberger and S. B. Treiman, Phys. Rev. {\bf 110}, 1178, 1478 (1958).

\item
M. A. Shifman, A. J. Vainstein, and V. I. Zakharov, Nucl. Phys. {\bf B147}, 385, 448
(1979).

\item
B. L. Ioffe and A. V. Smilga, Nucl. Phys. {\bf B232}, 109 (1984).

\item
V. M. Belyaev and Ya. I. Kogan, Pis'ma Zh. Eksp. Teor. Fiz. {\bf 37}, 611 (1983)
[JETP Lett. {\bf 37}, 730(1983)].

\item
C. B. Chiu, J. Pasupathy, and S. J. Wilson, Phys. Rev. {\bf D32}, 1786 (1985).

\item E. M. Henley, W-Y. P. Hwang, and L. S. Kisslinger, Phys. Rev. {\bf D46}, 431
(1992).

\item V. M. Belyaev, B. L. Ioffe, and Ya. I. Kogan, Phys. Lett. {\bf B151}, 290
(1985).

\item B. L. Ioffe, Nucl. Phys. {\bf 188B} (1981) 317; [E] {\bf B191}, 591(1981);
V. M. Belyaev and B. L. Ioffe, Zh. Eksp. Teor. Fiz. {\bf 83}, 876(1982)
[Sov. Phys. JETP {\bf 56}, 493(1982)]; B. L. Ioffe, Z. Phys. {\bf C18}, 67 (1983).

\item W-Y. P. Hwang, MIT-CTP-2498 and hep-ph/9601219, Z. Phys. {\bf C75}, 701 (1997).

\item S. Weinberg, Phys. Lett. {\bf B251} (1990) 288; Physica {\bf 96A}, 327(1979).

\item H. Georgi, {\it Weak Interactions and Modern Particle Theory}
(Benjamin/Cummings Publishing Co., Menlo Park, California, 1984); A. Manohar
and H. Georgi, Nucl. Phys. {\bf B234}, 189(1984).

\item W-Y. P. Hwang, unpublished; articles in preparation.

\end{enumerate}

\end{document}